\documentclass[aps,preprint,amsmath,amssymb]{revtex4}
\usepackage{psfig}% Include figure files
\usepackage{graphicx}%%%

\begin{document}

\title{Study of $B\to D^{**} \pi$ decays }

\author{\bf  Chuan-Hung Chen$^{a}$\footnote{Email:
physchen@mail.ncku.edu.tw}}

\vskip1.0cm

\affiliation{Department of Physics, National Cheng-Kung
University, Tainan 701, Taiwan }

\begin{abstract}
We investigate the production of the novel $P$-wave mesons
$D^{*}_{0}$ and $D^{\prime}_{1}\, (D_{1})$, identified as
$J^{P}=0^+$ and $1^+$, in heavy $B$ meson decays, respectively.
With the  heavy quark limit, we give our modelling wave functions
for the scalar meson $D^{*}_{0}$. Based on the assumptions of
color transparency and factorization theorem, we estimate the
branching ratios of $B\to D^{*}_{0} \pi$ decays in terms of the
obtained wave functions. Some remarks on $D^{(\prime)}_{1}$
productions are also presented.
\end{abstract}

\maketitle

\section{Introduction}

It is doubtless that quark model provides a successful method to
describe the hadron spectroscopy. For instance, based on the SU(3)
flavor symmetry, quark and anti-quark can comprise octet and
singlet states, called nonet together, such as the well known
pesudoscalar mesons of pion, kaon and eta
%belong to this nonet with the quantum number
with $J^{PC}=0^{-+}$. However, if we apply the same concept on the
light scalar mesons described by $0^{++}$, such as the nonet
%which is
composed of isoscalars $\sigma(600)$ and $f_{0}(980)$, isovector
$a_{0}(980)$ and isodoublet $\kappa$, there are some puzzles  (a)
why $a_{0}(980)$ and $f_0(980)$ are degenerate in masses and (b)
why the widths of $\sigma$ and $\kappa$ are broader than those of
$a_{0}(980)$ and $f_{0}(980)$ \cite{Cheng}. It is probable that
these scalar states consist of four-quark rather than two-quark
\cite{4q}.
% Besides the interpretation of $qq\bar{q}\bar{q}$
%four-quark states,
Moreover, the possibilities of $K\bar{K}$ molecular states,
gluonium states and scalar glueballs are also proposed. It is
clear that
% Hence,
the conclusion is still uncertain.

Now, the mysterious event happens not only in the light scalar
meson system, but also in the heavy $\bar{c}s$ one. Recently,
BABAR collaboration has observed one narrow state, denoted by
$D^{*}_{sJ}(2317)$, from a $D^{+}_{s} \pi^{0}$ mass distribution
\cite{Babar1}. Furthermore, the same state has been confirmed by
CLEO and a new state $D^{*}_{sJ}(2463)$ is also observed in the
$D^{*+}_{s} \pi^0$ final state \cite{CLEO}. Finally, BELLE
verifies the observations \cite{Belle1}. By the data analysis,
$D^{*}_{sJ}(2317)$ and $D^{*}_{sJ}(2463)$ are identified as
parity-even states with ${\bf 0}^{+}$ and ${\bf 1}^{+}$,
respectively.
% In BABAR, the mass of $D^{*}_{sJ}(2317)$ is
%determined to be $2316.8 \pm 0.4$ MeV and it's width ensure being
%less than $10$ MeV. However, CLEO not only confirms what BABAR
%observes but also measures the mass splitting of
%$D^{*}_{sJ}(2463)$ with respect to ordinary vector meson
%$D^{*}_{s}$ to be $351.6\pm 1.7 \pm 1.0$ MeV and finds the width
%being less than $7$ MeV.
 From the observations, the interesting
problem is that the states of $D^{*}_{sJ}(2317)$ and
$D^{*}_{sJ}(2463)$ cannot match with
%most
theoretical predictions \cite{Theory}, {\it i.e.}, the masses
(widths) are too low (narrow). To explain the discrepancy, either
the theoretical models have to be modified \cite{BEH} or the
observed states are the new composed states.
% It is clear that we need a new way to figure them out.
To satisfy the latter, many interesting solutions have been
suggested recently in
%by the authors of
Refs. \cite{CJ,BCL,BR,YH,Szczepaniak,CD,Chen-Li}.

In fact, before the BABAR's observation, CLEO \cite{CLEO-96} and
BELLE \cite{Belle-conf} measured the similar states in the
$\bar{c} q$ ($q=u,d$) system in heavy $B$ meson decays. With
two-quark picture, there are four parity-even (angular momentum
$\ell=1$) states  described by $J^P=0^{+}$, $1^{+}$, $1^{+}$ and
$2^{+}$, respectively. $J=j_{q}+S_{Q}$ is the total angular
momentum of the corresponding meson and consists of the angular
momentum of the light quark, $j_{q}$, and the spin of the heavy
quark, $S_{Q}$, where $j_{q}=S_{q}+\ell$ is combined by the spin
and orbital angular momenta of the light quark. In the literature,
they are usually labeled by $D^{*}_{0}$, $D^{\prime}_{1}$, $D_{1}$
and $D^{*}_{2}$, respectively. We also use $D^{**}$ to denote all
of them. The first two belong to $j_{q}=1/2$, while the last two
$j_{q}=3/2$. In the heavy quark limit, it is known that
$D^{*(\prime)}_{0(1)}$ and $D^{(*)}_{1(2)}$ decay only via $S$ and
$D$-wave, respectively.
%  only through $D$-wave.
Therefore, one expects that the widths of the former are much
broader than those of the latter, which
% The expectation
is consistent with the observations of CLEO and BELLE
\cite{CLEO-96,Belle-conf}. Even the BELLE's updated data
\cite{Belle-new} also show the same phenomenon. We now summarize
the results of CLEO and BELLE
%can be summarized
as follows: in CLEO, the masses (widths) of $P$-wave states are
given by $m_{D_{1}}(\Gamma_{D_{1}})=2422.0 \pm 2.1\;
(18.9^{+4.6}_{-3.5})$ and
$m_{D^{*}_{2}}(\Gamma_{D^{*}_{2}})=2458.9 \pm 2.0\; (23\pm 5)$
MeV, and the measured branching ratios (BRs) of $B$ decays are
given as
\begin{eqnarray}
BR(B^-\to D^{0}_{1} \pi^-) \times BR(D^{0}_{1} \to D^{*+}
\pi^-)=(7.8 \pm 1.9) \times 10^{-4}, \nonumber \\
BR(B^-\to D^{*0}_{2} \pi^-) \times BR(D^{*0}_{2} \to D^{*+}
\pi^-)=(4.2 \pm 1.7) \times 10^{-4}.
\end{eqnarray}
%%%%%%%%%%%%%%%%%%%%%%%%%%%%%%%%%%%%%%%%%%%%%%
In BELLE, the four states are all measured as
$m_{D^*_{0}}(\Gamma_{D^{*}_{0}})=2308 \pm 17 \pm 15 \pm 28\; (276
\pm 21 \pm 18 \pm 60)$,
$m_{D^{\prime}_{1}}(\Gamma_{D^{\prime}_{1}})=2427.0 \pm 26 \pm 20
\pm 15\; (384^{+107}_{-75}\pm 24 \pm 70)$,
$m_{D_{1}}(\Gamma_{D_{1}})=2421.4 \pm 1.5 \pm 0.4 \pm 0.8\; (23.7
\pm 2.7\pm 0.2 \pm 4.0)$, and
$m_{D^{*}_{2}}(\Gamma_{D^{*}_{2}})=2461.6 \pm 2.1 \pm 0.5 \pm
3.3\; (45.6 \pm 4.4 \pm 6.5 \pm 1.6)$ MeV, and the BRs  with
possible decaying chain being
\begin{eqnarray}
BR(B^-\to D^{*0}_{0} \pi^-) \times BR(D^{*0}_{0} \to D^{+}
\pi^-)=(6.1\pm0.6\pm 0.9 \pm 1.6) \times 10^{-4}, \nonumber
\\BR(B^-\to D^{'0}_{1} \pi^-) \times BR(D^{\prime0}_{1} \to D^{*+}
\pi^-)=(5.0\pm 0.4\pm 1.0 \pm 0.4) \times 10^{-4},\nonumber \\
BR(B^-\to D^{0}_{1} \pi^-) \times BR(D^{0}_{1} \to D^{*+}
\pi^-)=(6.8\pm0.7\pm 1.3 \pm 0.3) \times 10^{-4}, \nonumber \\
BR(B^-\to D^{*0}_{2} \pi^-) \times BR(D^{*0}_{2} \to D^{+}
\pi^-)=(3.4 \pm 0.3 \pm 0.6 \pm 0.4) \times 10^{-4},\nonumber \\
BR(B^-\to D^{*0}_{2} \pi^-) \times BR(D^{*0}_{2} \to D^{*+}
\pi^-)=(1.8 \pm 0.3 \pm 0.3 \pm 0.2) \times 10^{-4},
\label{belle-new}
\end{eqnarray}
respectively.

 Since the masses of $D^{*}_{sJ}$ are just below the
$D^{(*)}K$ threshold and the corresponding widths are around few
$10$ KeV \cite{YH}, both parity-even mesons could only decay
through isospin violating channels to $D\pi$ and $D^* \pi$. Due to
this reason, it becomes the main problem how to explain the low
masses and narrow widths for $D^{*}_{sJ}$ states. Unlike
$D^{*}_{sJ}$ cases, however, there are no any suppressions on
their decays to $D \pi$ or $D^* \pi$ although the measured masses
of $D^{**}$ are slightly different from the predictions of
theoretical models. It is believed that the properties of $D^{**}$
could be described by current theoretical models with some
improvements. If so, based on the concept of the normal quark
model, we could further understand the nature of $D^{**}$ in $B$
decays.

To handle the hadronic effects for $B\to D^{**} \pi$ decays, we
use the factorization formalism, called perturbative QCD (PQCD)
approach, which is based on factorization theorem and the
transition matrix element is described by the convolution of
hadron wave functions and the hard kernel \cite{LB,Li}. The wave
functions in principle can be extracted by experimental data or
determined by QCD sum rules or lattice calculations. The hard
kernel is related to the hard gluon exchange and high energetic
fermion propagator, which are all calculable perturbatively. At
the limit of heavy quark, in order to guarantee that color
transparency mechanism \cite{Bjorken} is satisfied, {\it i.e.}, no
soft gluon exchange between the final states, we need the
hierarchy of $m_{B}>>m_{D^{**}}>>\bar{\Lambda}$ with
$\bar{\Lambda}\sim m_{B}-m_{b}\sim m_{D^{**}}-m_{c}$ \cite{KLS}.
It has been shown by Ref. \cite{PQCD-Group} that with the same QCD
approach, the calculated results on $B\to D \pi$ decays are
consistent with the current observations \cite{PDG}. Consequently,
one expects that PQCD could be also applied to the $P$-wave meson
production in $B$ decaying processes. By the study, we should know
more properties related to $P$-wave mesons.

The paper is organized as follows. We investigate the
characteristic of $D^{**}$ and model their amplitude distributions
in Sec. \ref{ad}. In terms of PQCD approach, we derive the
factorization formulas for each $B\to D^*_{0} \pi$ decays in Sec.
\ref{ff}. We present our results
%with various values of the unknown parameters
in Sec. \ref{na}. Finally, we give a summary in Sec.
\ref{summary}.

\section{Decay constants and wave functions of $D^{**}$ \label{ad}}

In order to study the production of the scalar meson $D^{**}$ in
$B$ decays, we immediately have to face two questions.  The first
one is how to write down the hadronic structures and model the
wave functions of $D^{**}$, and the second is what the values of
decay constants of $D^{**}$ are. In PQCD, since the wave functions
belong to nonperturbative objects and are universal, we can
directly apply the wave functions of $B$ and $\pi$ mesons, which
have been discussed in other $B$-meson decays, such as $B\to \pi
\ell \nu$, $B\to \pi \pi$  etc. Therefore, the hadronic structures
of $B$ and $\pi$
mesons can be described by \cite{KLS-PRD,Ball}%%%%%%%%%%%%%%%
\begin{eqnarray}
\langle 0|\bar{b}(0)_{j} d(z)_{l}|B,p_{1}\rangle
&=&\frac{1}{\sqrt{2N_{c}}}\int^{1}_{0} dx e^{-ixp_{1}\cdot z}
 \Big\{ [ \slash \hspace{-0.2cm}
p_{1} +m_{B} ]_{lj}\gamma_{5} \Phi_{B}(x)\Big\}, \nonumber\\
%\end{eqnarray}
%%%%%%%%%%%%%%%%%%%%%%%%%%%%%%%%%%%%%%%%%%%%%%%%%%%
%\begin{eqnarray}
\langle 0|\bar{u}(0)_{j}
d(z)_{l}|\pi,p_{3}\rangle&=&\frac{1}{\sqrt{2N_c}}\int^{1}_{0} dx
e^{-ixp_{3}\cdot z} \Big[[\slash \hspace{-0.2
cm}{p_{3}}]_{lj}\Phi_{\pi}(x) \nonumber \\
&& +m^{0}_{\pi}[I]_{lj} \Phi_{\pi}^{p}(x)+m^{0}_{\pi}[\slash
\hspace{-0.2 cm}{n_{-}} \slash \hspace{-0.2
cm}{n_{+}}-I]_{lj}\Phi^{\sigma}_{\pi}(x)\Big], \label{bpi}
\end{eqnarray}
%%%%%%%%%%%%%%%%%%%%%%%%%%
where $n_{-}=(0,1,\vec{0}_{T})$, $n_{+}=(1,0,\vec{0}_{T})$ and $x$
is the momentum fraction of the light parton inside the
corresponding meson. $\Phi_{\pi}$ and $\Phi^{p(\sigma)}_{\pi}$ are
the twist-2 and twist-3 pion wave functions,  related to the
distribution amplitude of nonlocal operator $\bar{u} \gamma_{5}
\gamma_{\mu} d(z)$ and associated with $\bar{u} \gamma_{5} d(z)$
($\bar{u} \gamma_{5} \sigma_{\mu\nu} d(z)$), respectively. We note
that $m^{0}_{\pi}$ is the so-called chiral symmetry breaking
parameter and is equivalent to $m^{2}_{\pi}/(m_{u}+m_{d})$.

To determine the structures and distribution amplitudes of
$D^{**}$, we need to know their properties. For simplicity, we
only concentrate the discussion on the scalar meson of
$D^{*}_{0}$. The similar analysis can be applied to other charmed
$P$-wave mesons. As usual, the decay constant of $D^{*}_{0}$ is
defined as
\begin{equation}
\langle 0 |\bar{d}\; c|D^{*}_{0},p_{2}
\rangle=m_{D^{*}_{0}}\tilde{f}_{D^{*}_{0}}.\label{sc}
\end{equation}
By using the equation of motion, we  obtain another identity
\begin{equation}
\langle 0 |\bar{d}\, \gamma_{\mu}\, c|D^{*}_{0},p_{2}
\rangle=f_{D^{*}_{0}}  p_{2\mu}, \label{vc}
\end{equation}
with $f_{D^{*}_{0}}= \tilde{f}_{D^{*}_{0}} (m_c -m_d
)/m_{D^{*}_{0}}$, in which $m_{c(d)}$ are the current quark mass
of $c(d)$-quark.
 From the above equation, we see that if the considering case is light scalar meson,
the corresponding transition matrix element will become small.
This is the reason why the decay constant of the light scalar
meson for vector current is small.
%Hence,
In order to satisfy the conditions of Eqs. (\ref{sc}) and
(\ref{vc}), the hadronic structure of $D^{*}_{0}$ is adopted to be
\begin{eqnarray}
\langle D^{*}_{0},p_{2}|\bar{d}(0)_{j} c(z)_{l}|0\rangle
&=&\frac{1}{\sqrt{2N_{c}}}\int^{1}_{0} dx e^{ixp_{2}\cdot z}
 \Big\{[ \slash \hspace{-0.2cm}
p_{2}]_{lj} \Phi_{D1}(x) + m_{D^{*}_{0}} [I ]_{lj}\Phi_{D2}(x)
\Big\}, \label{sdwf}
\end{eqnarray}
%%%%%%%%%%%%%%%%%%
with the normalizations,
\begin{eqnarray*}
\int^{1}_{0} \Phi_{D1}(x)&=&\frac{f_{D^{*}_{0}}}{2\sqrt{2N_{c}}},
\ \ \ \int^{1}_{0}
\Phi_{D2}(x)=\frac{\tilde{f}_{D^{*}_{0}}}{2\sqrt{2N_{c}}}.
\end{eqnarray*}

The value of decay constant $\tilde{f}_{D^{*}_{0}}$ is the crucial
part for concerning whether $D^{*}_{0}$ production is interesting
or not. To estimate the magnitude of $\tilde{f}_{D^{*}_{0}}$, we
need the help with the scalar meson $K^*_{0}(1420)$, for which the
decay constat has already been estimated in Ref. \cite{Maltman}.
As mentioned early, the scalar meson generally satisfies the
identity
\begin{eqnarray*}
(m_{q_{1}}-m_{q_{2}})\langle 0| \bar{q}_{2} q_{1} | S \rangle=
 f_{S} m^{2}_{S},
\end{eqnarray*}
where $m_{q_{i}}$, $m_{S}$ and $f_{S}$ are the current quark mass,
the $S$-meson mass and its decay constant of vector current,
respectively. If we assume $\langle 0| \bar{d}\, s | K^*_{0}(1420)
\rangle \approx \langle 0| \bar{d}\, c | D^{*}_{0} \rangle$, from
the above equation, we can obtain $f_{D^*_{0}}=f_{K^*_{0}} \cdot
m^{2}_{K^*_{0}}/m^{2}_{D^*_{0}}\cdot (m_{c}-m_{d})/(m_{s}-m_{d})$.
With the values of $f_{K^*_{0}}\sim 34$ MeV \cite{Maltman},
$m_{c}=1.5$ GeV, $m_{s}=150$ and $m_{d}=8.7$ MeV, we get
$f_{D^*_{0}}\approx 130$ MeV. This value is close to the result in
Ref. \cite{VD}, calculated by relativistic quark model. Finally,
from Eq. (\ref{vc}) we have
$\tilde{f}_{D^{*}_{0}}=m_{D^*_{0}}/(m_{c}-m_{d}) \cdot
f_{D^*_{0}}\approx 200$ MeV. It is known that $K^*_{0}$ is
composed of a two-quark state. Thus, it is interesting to have the
similar decay constants between the scalar $D^{*}_{0}$ and
pseudoscalar $D_{s}$.

To obtain the shapes of $D^{*}_{0}$ wave functions qualitatively,
we need to employ the concept of the heavy quark limit.  According
to Eq. (\ref{sdwf}), we see that $\Phi_{D1}(x)$ is the
distribution amplitude of the nonlocal operator $\bar{d}
\gamma_{\mu} c (z)$ while $\Phi_{D2}(x)$ is associated with
$\bar{d}\, c (z)$. By the equation of motion, we straightforwardly
find that the difference between $\Phi_{D1}(x)$ and $\Phi_{D2}(x)$
is order of
$\bar{\Lambda}/m_{D^{*}_{0}}\sim(m_{D^*_{0}}-m_{c})/m_{D^*_{0}}$.
Hence, if we set $m_{D^*_{0}} \sim m_{c}$, we can get the
information of $\Phi_{D1}(x)\sim \Phi_{D2}(x)$. Furthermore, in
order to satisfy the identities of decay constants defined by Eqs.
(\ref{sc}) and (\ref{vc}), the simplest forms for both wave
functions can be modelled by $\Phi_{Di}\propto x(1-x)+a_{i}
x(1-x)(1-2x)$ in which $a_{i}$ are free parameters. Since the
second term is antisymmetric while $x$ is replaced by $1-x$, we
can easily conclude that this term will not change the
normalization of the wave function. Therefore, we could use it to
control the shapes of the wave function. It is worth to mention
that since we consider $D^{*}_{0}$ to be a $P$-wave state,
% it could be thought that
the size of $D^{*}_{0}$ is believed to be bigger than that of
particle in the $S$-wave state. In order to avoid that $D^{*}_{0}$
becomes oversize such that the mechanism of color transparency is
breakdown, like the $b$-dependence on the wave function of the $B$
meson, in which $b$ is the conjugate variable of the parton
transverse momentum, we also introduce the intrinsic
$b$-dependence on $D^{*}_{0}$.
% for controlling its size.
To satisfy Eqs. (\ref{sc}) and (\ref{vc}), the final simplest
shapes of the wave functions are expressed as
%%%%%%%%%%%%%%%
\begin{eqnarray}
\Phi_{D1}(x,b)&=&\frac{\tilde{f}_{D^{*}_{0}}}{2\sqrt{2N_{c}}}\Big\{
6x(1-x)\Big[ {m_{c}-m_{d} \over m_{D^{*}_{0}}} + a_{D^{*}_{0}}
(1-2x)\Big]\Big\} \exp\left[-\frac{\omega^{2}_{D^{*}_{0}}
b^2}{2}\right]
, \nonumber \\
\Phi_{D2}(x,b)&=&\frac{\tilde{f}_{D^{*}_{0}}}{2\sqrt{2N_{c}}}
\Big\{6x(1-x)[1 + b_{D^{*}_{0}}
(1-2x)]\Big\}\exp\left[-\frac{\omega^{2}_{D^{*}_{0}}
b^2}{2}\right] ,
\end{eqnarray}
where $\omega_{D^{*}_{0}}$, $a_{D^{*}_{0}}$ and $b_{D^{*}_{0}}$
are the unknown parameters. Although $b_{D^{*}_{0}}$ is a free
parameter, it can be chosen such that the $D^{*}_{0}$ meson wave
function has the maximum at $x\approx
(m_{D^{*}_{0}}-m_{c})/m_{D^{*}_{0}}\sim 0.35$ for $m_{c}=1.5$ GeV.
As to the value of $a_{D^{*}_{0}}$, we refer to the case of
$K^*_{0}(1410)$ \cite{DH}. By assuming that $
a_{D^*_{0}}{m_{D^*_{0}}}/(m_{c}-m_{d}) \sim
a_{K^*_{0}}{m_{K^*_{0}}}/(m_{s}-m_{d}) \approx 75/f_{K^*_{0}}$,
the order of magnitude of $a_{D^*_{0}}$ is estimated to be around
$1.2$.
%% Nevertheless, in our
%%calculations, we still take them as the free parameters.

\section{Factorization formulas \label{ff}}

Since the considered decays $B\to D^{**} \pi$ correspond to the
$b\to c\, \bar{u}\, d$ transition, we describe the effective
Hamiltonian as
\begin{eqnarray}
H_{{\rm eff}}&=&\frac{G_{F}}{\sqrt{2}}V_{c}\left[ C_{1}(\mu
)\bar{d}_{\alpha} u_{\beta} \bar{c}_{\beta} b_{\alpha} +C_{2}(\mu
) \bar{d}_{\alpha} u_{\alpha} \bar{c}_{\beta} b_{\beta} \right],
 \label{eff}
\end{eqnarray}
where $\bar{q}_{\alpha} q_{\beta}=\bar{q}_{\alpha} \gamma_{\mu}
(1-\gamma_{5}) q_{\beta}$, $\alpha(\beta)$ are the color indices,
$V_{c}=V_{ud}^{*}V_{cb}$ is the product of the CKM matrix elements
\cite{CKM}, and $C_{1,2}(\mu )$ are the Wilson coefficients (WCs)
\cite{BBL}. With the light-cone coordinate, the momenta of various
mesons and the light valence quarks inside the corresponding
mesons are assigned as: $p_{1}=m_{B}/\sqrt{2}(1,1,\vec{0}_{T})$,
$k_{1}=m_{B}/\sqrt{2}(x_{1},0,\vec{k}_{1T})$;
$p_{2}=m_{B}/\sqrt{2}(1,r^{2}_{2},\vec{0}_{T})$,
$k_{2}=m_{B}/\sqrt{2}(x_{2},0,\vec{k}_{2T})$;
$p_{3}=m_{B}/\sqrt{2}(0,1-r^{2}_{2},\vec{0}_{T})$,
$k_{3}=m_{B}/\sqrt{2}(0,(1-r^{2}_{2})x_{3},\vec{k}_{3T})$, with
$r_{2}=m_{D^{*}_{dJ}}/m_{B}$. As usual, we use
\begin{eqnarray}
\Gamma=\frac{G^{2}P_{c}m^{2}_{B}}{16\pi }|V_{c}|^{2} |{\cal
A}|^{2} \label{gamma}
\end{eqnarray}
to describe the decay rates of $B\to D^{**} \pi$, in which
$P_c\equiv |p_{2z}|=|p_{3z}|\approx m_{B}(1-r^{2}_{2})/2$ is the
momentum of the outgoing meson, ${\cal A}$ is the decay amplitude
and its value depends on QCD approaches. Since the hadronic
structures of the tensor meson haven't been derived yet and so far
they are not definite,
%to be more careful,
we study the problem elsewhere.
%as our future work.
Although $D^{\prime}_{1}$ and $D_{1}$ are the vector mesons and
carry the spin degrees of freedom, only longitudinal polarization
has the contribution since one of the final states is a
pseudoscalar. Therefore, the deriving formulas for $B \to
D^{*}_{0} \pi$ are also proper to the final states with one vector
and one pseudoscalar mesons. In this paper, we only concentrate on
the production of $D^*_{0}$.

In terms of the effective interactions, we see that different
decaying processes involve different topologies. To be more clear,
in the following we analyze each of $B\to D^{*}_{0} \pi$ decays
separately.

${\it 1.\ B_{d}\to D^{*-}_{0} \pi^+\ decay:}$
%%%%%%%%%%%%%%Figure %%%%%%%%%%%%%%%%%
\begin{figure}[phtb]\begin{center}
\includegraphics*[width=2.8
in]{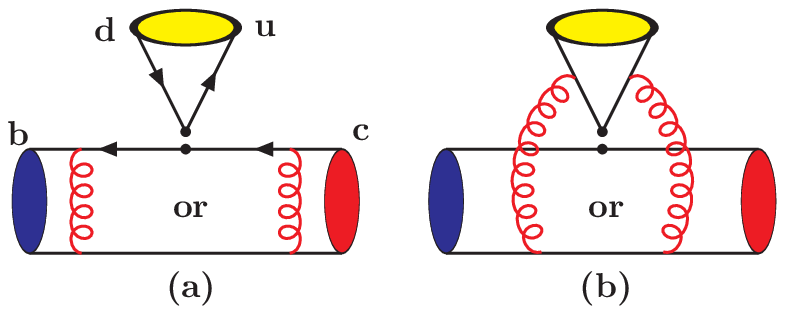}\hspace{0.5cm}\includegraphics*[width=2.8
in]{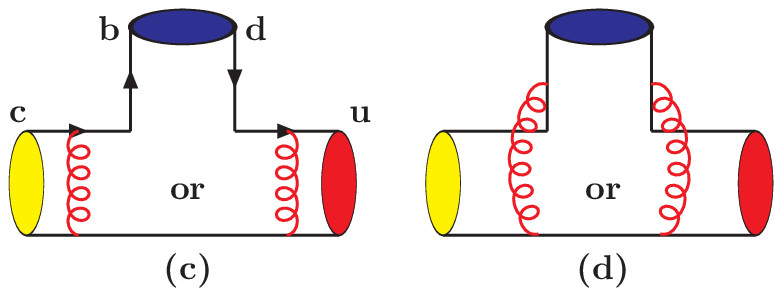}% Here is how to import EPS art
\caption{ The topologies (a)[(c)] factorizable  emission
[annihilation] and (b)[(d)] nonfactorizable effects for the decays
$B_{d}\to D^{*-}_{0} \pi^{+}$.} \label{ca_fig} \end{center}
\end{figure}
%%%%%%%%%%%%%%%%%%%%%%%%%%%%%%%%%%%%%%%%%%%%%%%

There are two topologies in this decay, emission and annihilation
diagrams. The former is color-allowed but the latter belongs to
color-suppressed. The corresponding flavor diagrams are
illustrated by Fig. \ref{ca_fig}. Hence, the decay amplitude of
$B_{d} \to D^{*-}_{0} \pi^{+}$ can be expressed by
\begin{eqnarray}
{\cal A}(B_{d} \to D^{*-}_{0} \pi^{+})=f_{\pi} F_{BD^{*}_{0}} +
M_{BD^{*}_{0}}+f_{B} F_{a}+M_{a},
\end{eqnarray}
where $f_{\pi}$ and $f_{B}$ are the decay constants of $\pi$ and
$B$ mesons and the related contributions are the factorizable
emission and annihilation topologies, respectively. The remains
denote nonfactorizable contributions. With factorization theorem
and  hadronic structures of Eqs. (\ref{bpi}) and (\ref{sdwf}), the
hard amplitudes $\{F\}$ and $\{ M \}$ are formulated as
%%%%%%%%%%%%%%%%%%%%%%%%%%%%%%%%%%%%%%%%%%%%%%%%%%%%%
\begin{eqnarray}
F_{BD^{*}_{0}}&=&8 \pi C_{F} m^{2}_{B} \int_{0}^{1} dx_{1} dx_{2}
\int_{0}^{\infty} b_{1} db_{1} b_{2} db_{2} \Phi_{B}(x_{1},b_{1})
\nonumber\\
&& \times \Big\{ \Phi_{D1}(x_{2},b_{2}){\cal
E}_{D^{*}_{0}}^{1}(t^{(1)}_{e})h_{e}(x_{1},x_{2},b_{1},b_{2})\nonumber
\\&&+ (r_{c}\Phi_{D1}(x_{2})+2r_{2}\Phi_{D2}(x_{2})){\cal
E}_{D^{*}_{0}}^{2}(t^{(2)}_{e})
h_{e}(x_{2},x_{1},b_{2},b_{1})\Big\}
\label{fe1}\\
%\end{eqnarray}
%%%%%%%%%%%%%%%%%%%%%%%%%%%%%%%%%%%%%%%%
%\begin{eqnarray}
M_{BD^{*}_{0}}&=&16\pi C_F m_{B}^{2}\sqrt{2N_{c}} \int_{0}^{1}
dx_{1} dx_{2}dx_{3} \int_{0}^{\infty} b_{1} db_{1} b_{3} db_{3}
\Phi_{B}(x_{1},b_{1})
 \nonumber \\
&& \times\Phi_{\pi}(x_{3})\Phi_{D1}(x_{2},b_{1}) \Big\{
-(x_{2}+x_{3}) {\cal E}_{D^{*}_{0}}^{\prime
1}(t^{(1)}_{d})h^{(1)}_{d}(\{x\},b_{1}, b_{3}) \nonumber \\
&&+ (1-x_{3}) {\cal E}_{D^{*}_{0}}^{\prime
2}(t^{(2)}_{d})h^{(2)}_{d}(\{x\},b_{1}, b_{3})
\Big\},\label{nonfe1}
\end{eqnarray}
%%%%%%%%%%%%%%%%%%%%%%%%%%%%%%%%%%%%%%%%
\begin{eqnarray}
F_{a}&=&8 \pi C_{F} m^{2}_{B} \int_{0}^{1} dx_{2} dx_{3}
\int_{0}^{\infty} b_{2} db_{2}
db_{3}\Phi_{D1}(x_{2},b_{2})\Phi_{\pi}(\zeta) \nonumber \\&&
\Big\{- x_{3} {\cal E}_{a}^{1}(t^{(1)}_{a})
h_{a}(x_{2},x_{3}\eta_{3},b_{2},b_{3})+ x_{2}{\cal
E}_{a}^{2}(t^{(2)}_{a}) h_{a}(x_{3},x_{2}\eta_{3},b_{3},b_{2})
\Big\}, \label{fa}\\
%\end{eqnarray}
%%%%%%%%%%%%%%%%%%%%%%%%%%%%%%%%%%%%%%%%
%\begin{eqnarray}
M_{a}&=&16\pi C_F m_{B}^{2}\sqrt{2N_{c}} \int_{0}^{1} d[x]
\int_{0}^{\infty} [b]d[b]
\Phi_{B}(x_{1},b_{1})\Phi_{D1}(x_{2},b_{2})\Phi_{\pi}(\zeta)
 \nonumber \\
&& \times \Big\{x_{3} {\cal E}_{a}^{\prime
1}(t^{(1)}_{f})h^{(1)}_{f}(\{x\},b_{1},b_{2})-x_{2} {\cal
E}_{a}^{\prime
2}(t^{(2)}_{f})h^{(2)}_{f}(\{x\},b_{1},b_{2})\Big\}.\label{nonfa}
\end{eqnarray}
%%%%%%%%%%%%%%%%%%%%%%%%%%%%%%%%%%%%%%%%
The hard functions $h_{e(d,a,f)}$, related to the propagators of
exchange hard gluon and internal quark, are described by
\begin{eqnarray*}
h_{e}(x_1,x_2,b_1,b_2)&=&S_{t}(x_{2})K_{0}\left(\sqrt{x_1x_2}m_Bb_1\right)
\nonumber \\
& &\times \left[\theta(b_1-b_2)K_0\left(\sqrt{x_2}m_B
b_1\right)I_0\left(\sqrt{x_2}m_Bb_2\right)\right.
\nonumber \\
& &\left.+\theta(b_2-b_1)K_0\left(\sqrt{x_2}m_Bb_2\right)
I_0\left(\sqrt{x_2}m_Bb_1\right)\right]\;, \label{dh}
\end{eqnarray*}
%%%%%%%%%%%%%%%%%%%%%%%%%
\begin{eqnarray*}
h^{(j)}_{d}(x_{1},x_{2},x_{3}, b_{1},b_{2})&=&
\left[\theta(b_1-b_2)K_0\left(\sqrt{x_{1}x_{2}} m_B
b_1\right)I_0\left(\sqrt{x_{1}x_{2}} m_Bb_2\right)\right. \nonumber \\
& &\quad \left. +\theta(b_2-b_1)K_0\left(\sqrt{x_{1}x_{2}}m_B
b_2\right) I_0\left(\sqrt{x_{1}x_{2}}m_B b_1\right)\right]
 \nonumber \\
&  & \times \left( \begin{array}{cc}
 K_{0}(D_{j}m_Bb_{2}) &  \mbox{for $D^2_{j} \geq 0$}  \\
 \frac{i\pi}{2} H_{0}^{(1)}(\sqrt{|D_{j}^2|}m_Bb_{2})  &
 \mbox{for $D^2_{j} \leq 0$}
  \end{array} \right),
\label{hjd}\end{eqnarray*}
%%%%%%%%%%%%%%%%%%%%%%%%%%%%%%%%%%
\begin{eqnarray*}
h_a(x_2,x_3,b_2,b_3)&=&S_{t}(x_{3})\left(i\frac{\pi}{2}\right)^2
H_0^{(1)}\left(\sqrt{x_2x_3}m_Bb_2\right)
\nonumber \\
& &\times\left[\theta(b_2-b_3)
H_0^{(1)}\left(\sqrt{x_3}m_Bb_2\right)
J_0\left(\sqrt{x_3}m_Bb_3\right)\right.
\nonumber \\
& &\left.+\theta(b_3-b_2)H_0^{(1)}\left(\sqrt{x_3}m_Bb_3\right)
J_0\left(\sqrt{x_3}m_Bb_2\right)\right],
\end{eqnarray*}
%%%%%%%%%%%%%%%%%%%%%%%%%%%%%%%%
\begin{eqnarray*}
h^{(j)}_f(\{x\},b_{1},b_{2})&=& i\frac{\pi}{2}
\left[\theta(b_1-b_2)H_0^{(1)}\left(\sqrt{x_{2}x_{3}\eta_{3}}m_B
b_1\right)J_0\left(\sqrt{x_{2}x_{3}\eta_{3}}m_Bb_2\right)\right. \nonumber \\
& &\quad\left.
+\theta(b_2-b_1)H_0^{(1)}\left(\sqrt{x_{2}x_{3}\eta_{3}}m_B
b_2\right)
J_0\left(\sqrt{x_{2}x_{3}\eta_{3}}m_B b_1\right)\right]\;  \nonumber \\
&  & \times \left( \begin{array}{cc}
 K_{0}(F_{j}m_Bb_{1}) &  \mbox{for $F^2_{j} \geq 0$}  \\
 \frac{i\pi}{2} H_{0}^{(1)}(\sqrt{|F_{j}^2|}m_Bb_{1})  &
 \mbox{for $F^2_{j} \leq 0$}
  \end{array} \right),
\end{eqnarray*}
with $D^{2}_{1}=x_{1}x_{2}-x_{2}x_{3}\eta_{3}$,
$D^{2}_{2}=x_{1}x_{2}-x_{2}(1-x_{3})\eta_{3}$,
$F^{2}_{1}=(x_{1}-x_{2})x_{3}\eta_{3}$,
$F^{2}_{2}=x_{1}+x_{2}+(1-x_{1}-x_{2})x_{3}\eta_{3}$,
$\eta_{3}=(1-r^2_{2})$ and $r_{c}=m_{c}/m_{B}$.
%%%%%%%%%%%%%%%%%%%%%%%%%%%%%%%%%%%%%%%%
The threshold resummation effect is expressed to be
$S_t(x)=2^{1+2c} \cdot \Gamma(3/2+c)
[x(1-x)]^c/(\sqrt{\pi}\Gamma(1+c))$, with $c\approx 0.35$
\cite{TLS}. The evolution factors ${\cal E}_{D^{*}_{0}}^{i} ({\cal
E}_{D^{*}_{0}}^{\prime i})$ and ${\cal E}_{a}^{i} ({\cal
E}_{a}^{\prime i})$ are defined by
\begin{eqnarray*}
{\cal E}^{i}_{D^{*}_{0}}(t^{(i)}_{e})&=&\Big(C_{2}(t^{(i)}_{e})
+\frac{C_{1}(t^{(i)}_{e})}{N_{c}}\Big)\alpha_{s}(t^{(i)}_{e})
\exp[-S_{B}-S_{D^{*}_{0}}], \nonumber \\%%%%%%%%%%%%%%%%%%%%%%%%%%%
{\cal E}^{\prime
i}_{D^{*}_{0}}(t^{(i)}_{d})&=&\frac{C_{1}(t^{(i)}_{d})}{N_{c}}\alpha_{s}(t^{(i)}_{d})
\exp[-S_{B}-S_{D^{*}_{0}}-S_{\pi}]_{b_{2}=b_{1}}, \nonumber
\\%%%%%%%%%%%%%%%%%%%%%%%%%%%%%
{\cal E}^{i}_{a}(t^{(i)}_{a})&=&\Big(C_{1}(t^{(i)}_{a})
+\frac{C_{2}(t^{(i)}_{a})}{N_{c}}\Big)\alpha_{s}(t^{(i)}_{a})
\exp[-S_{D^{*}_{0}}-S_{\pi}]_{b_{3}=b_{2}}, \nonumber \\%%%%%%%%%%%%%%%%%%%%%%%%
{\cal E}^{\prime
i}_{a}(t^{(i)}_{f})&=&\frac{C_{2}(t^{(i)}_{f})}{N_{c}}\alpha_{s}(t^{(i)}_{f})
\exp[-S_{B}-S_{D^{*}_{0}}-S_{\pi}]_{b_{3}=b_{2}},
\end{eqnarray*}
where the exponents $S_{M}$ ($M=B, D^{*}_{0}, \pi$) are the
Sudakov factors. From above equations, we see clearly that the
emission contributions are color-allowed and dictated by effective
coupling of $C_{2}+C_{1}/N_{c}$, while the annihilation
contributions are color-suppressed and governed by
$C_{1}+C_{2}/N_{c}$. $t^{(i)}_{e,d,a,f}$ denote the hard scales of
the involving diagrams which are expected to be of ${\cal
O}(\sqrt{\bar{\Lambda} m^{2}_{B}})\sim 1.6$ GeV in average and the
criteria to determine them are adopted to be
%%%%%%%%%%%%%%%%%%%%%%%%%%%%%%%%%%%%%%%%%%
\begin{eqnarray}
t_e^{(1)}&=&{\rm max}(\sqrt{x_2}m_B,1/b_1,1/b_2), \;\;\;
t_e^{(2)}={\rm max}(\sqrt{x_1}m_B,1/b_1,1/b_2), \nonumber
\\
%%%%%%%%%%%%%%
t_{d}^{(j)}&=&{\rm max}(\sqrt{x_{1} x_{2}}m_{B},
\sqrt{D^{2}_{j}}m_B,1/b_1,1/b_3), \nonumber \\
%%%%%%%%%%%%%%%%
t_a^{(1)}&=&{\rm max}(\sqrt{x_{3}}m_B,1/b_2,1/b_3), \;\;\;
t_a^{(2)}={\rm max}(\sqrt{x_{2}}m_B,1/b_2,1/b_3),\nonumber \\
%%%%%%%%%%%%%%%%%%%%
t_f^{(j)}&=&{\rm
max}(\sqrt{x_{2}x_{3}\eta_{3}}m_B,\sqrt{F_j^2}m_B,1/b_1,1/b_2).
\label{scale}
\end{eqnarray}
Since we deal with the hadronic effects of the $B$ decay by
considering six-quark simultaneously, at lowest order in strong
interaction, besides the renormalization group (RG) running from
$m_{W}$ to $m_{B}$ scales in the $\mu$-scale dependence of WCs, we
still need to consider the running from $m_{B}$ scale to the hard
scale $t^{(i)}_{e,d,a,f}$ which indeed dictate the scale of the
$B$ meson decay. Hence, in our consideration, the hard scales for
WCs are determined by Eq. (\ref{scale}) rather than at $m_{B}$ or
$m_{B}/2$ scale. In the formulations of Eqs.
(\ref{fe1})$-$(\ref{nonfa}), we have dropped the terms related to
$r^{2}_{2}$ ($r_{c}$ and $r^{2}_{2}$) for the right-handed
(left-handed) gluon exchange of Fig. \ref{ca_fig}. Compared to
leading power, which isn't suppressed by $1/m_{B}$, they all
belong to higher power effects.

${\it 2.\ B_{d}\to \bar{D}^{*0}_{0} \pi^0\ decay:}$
%%%%%%%%%%%%%%Figure %%%%%%%%%%%%%%%%%
\begin{figure}[phtb]\begin{center}
\includegraphics*[width=2.8
in]{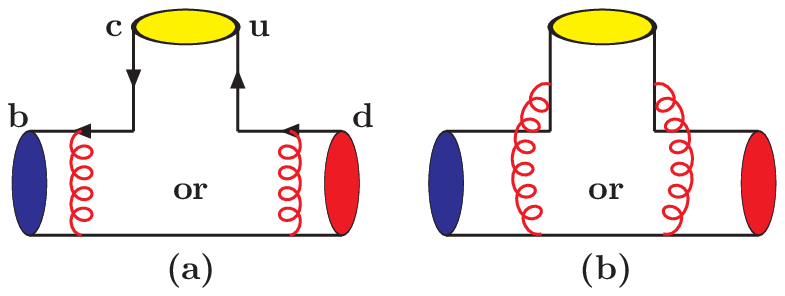}% Here is how to import EPS art
\caption{ The topologies (a) factorizable  emission and (b)
nonfactorizable effects for the decays $B_{d}\to \bar{D}^{*0}_{0}
\pi^{0}$.} \label{cs_fig} \end{center}
\end{figure}
%%%%%%%%%%%%%%%%%%%%%%%%%%%%%%%%%%%%%%%%%%%%%%%

In this decay, the involving annihilation contributions are the
same as the decay $B_{d}\to D^{*-}_{0} \pi^+$ but the emission
topologies become color-suppressed, illustrated by Fig.
\ref{cs_fig}. Due to the neutral pion meson being described by
$(\bar{u}u-\bar{d}d)/\sqrt{2}$, the sign of emission topologies is
opposite to that of annihilation topologies. Therefore, the decay
amplitude is written as
\begin{eqnarray}
{\cal A}(B_{d} \to \bar{D}^{*0}_{0} \pi^{0})=\frac{1}{\sqrt{2}}
\Big[- f_{D^{*}_{0}} F_{B\pi} - M_{B\pi}+f_{B} F_{a}+M_{a}\Big],
\label{nnamp}
\end{eqnarray}
With the same approach and power counting for the $B_{d}\to
D^{*-}_{0} \pi^+$ decay, the relevant hard amplitudes $F_{B\pi}
(M_{B\pi})$ can be derived as
\begin{eqnarray}
F_{B\pi}&=&8 \pi C_{F} m^{2}_{B} \int_{0}^{1} dx_{1} dx_{3}
\int_{0}^{\infty} b_{1}db_{1} b_{3}db_{3} \Phi_{B}(x_{1},b_{1})
\nonumber\\
&& \times \Big\{
\Big[(1+x_{3})\Phi_{\pi}(x_{3})+r_{\pi}(1-2x_{3})(\Phi^{p}_{\pi}(x_{3})
+\Phi^{\sigma}_{\pi}(x_{3})) \Big] {\cal
E}_{\pi}^{3}(t^{(3)}_{e}) \nonumber \\
&& \times   h_{e}(x_{1},x_{3}\eta_{3},b_{1},b_{3})
+2r_{\pi}\Phi^{p}_{\pi}(x_{3}){\cal E}_{\pi}^{4}(t^{(4)}_{e})
h_{e}(x_{3},x_{1}\eta_{3},b_{3},b_{1})
 \Big\},
\label{fe2}\\
%\end{eqnarray}
%%%%%%%%%%%%%%%%%%%%%%%%%%%%%%%%%%%%%%%%
%\begin{eqnarray}
M_{B\pi}&=&16\pi C_F m_{B}^{2}\sqrt{2N_{c}} \int_{0}^{1} dx_{1}
dx_{2} dx_{3} \int_{0}^{\infty} b_{1}db_{1} b_{2} db_{2}
\Phi_{B}(x_{1},b_{1})\Phi_{D1}(x_{2},b_{2})
 \nonumber \\
&& \times \Big\{ \Big[ -(x_{2}+x_{3})\Phi_{\pi}(x_{3})
+r_{\pi}x_{3}(\Phi^{p}_{\pi}(x_{3}) + \Phi^{\sigma}_{\pi}(x_{3}) )
\Big]\nonumber \\
&& \times {\cal E}_{\pi}^{\prime
3}(t^{(3)}_{d})h^{(3)}_{d}(x_{1},x_{3}\eta_{3},x_{2},b_{1}, b_{2}) \nonumber \\
&&+\Big[(1-x_{2})\Phi_{\pi}(x_{3})-r_{\pi}x_{3}
(\Phi^{p}_{\pi}(x_{3}) + \Phi^{\sigma}_{\pi}(x_{3}) )
\Big]\nonumber \\
&&\times {\cal E}_{\pi}^{\prime 4}(t^{(4)}_{d}) h^{(4)}_{d}(x_{1},
x_{3}\eta_{3},x_{2} ,b_{1}, b_{2}) \Big\}. \label{nonfe2}
\end{eqnarray}
The evolution factors ${\cal E}_{\pi}^{i} ({\cal E}_{\pi}^{\prime
i})$ are defined by
\begin{eqnarray*}
{\cal E}^{i}_{\pi}(t^{(i)}_{e})&=&\Big(C_{1}(t^{(i)}_{e})
+\frac{C_{2}(t^{(i)}_{e})}{N_{c}}\Big)\alpha_{s}(t^{(i)}_{e})
\exp[-S_{B}-S_{\pi}] , \nonumber \\
{\cal E}^{\prime
i}_{\pi}(t^{(i)}_{d})&=&\frac{C_{2}(t^{(i)}_{d})}{N_{c}}\alpha_{s}(t^{(i)}_{d})
\exp[-S_{B}-S_{D^{*}_{0}}-S_{\pi}]_{b_{3}=b_{1}}.
\end{eqnarray*}
%%%%%%%%%%%%%%%%%%%%%%%%%%
 From above equations, due to the appearance of
$C_{1}+C_{2}/N_{c}$, we know that $B_{d}\to \bar{D}^{*0}_{0}
\pi^0$ is color-suppressed process.  We note that although
nonfactorizable effects are also color-suppressed, since
$C_{2}/N_{c}$ could be larger than $C_{1}+C_{2}/N_{c}$, the
nonfactorizable effects play a important role in this kind of
color-suppressed processes. In fact, the same thing also happens
in the decay $B_{d}\to \bar{D}^{0} \pi^0$ with $\bar{D}^{0}$ being
charmed pseudoscalar \cite{PQCD-Group}. The hard scales are
determined by
\begin{eqnarray*}
t_e^{(3)}&=&{\rm max}(\sqrt{x_3 \eta_{3}}m_B,1/b_1,1/b_3)\;,
\;\;\; t_e^{(4)}={\rm max}(\sqrt{x_1 \eta_{3}}m_B,1/b_1,1/b_3)\;,
\nonumber \\
t_{d}^{(3)}&=&{\rm max}(\sqrt{x_{1} x_{3}\eta_{3}}m_{B},
\sqrt{D^{2}_{3}}m_B,1/b_1,1/b_2),\nonumber \\
 t_{d}^{(4)}&=&{\rm
max}(\sqrt{x_{1}
x_{3}\eta_{3}}m_{B},\sqrt{D^{2}_{4}}m_B,1/b_1,1/b_2),
\end{eqnarray*}
with $D^{2}_{3}=(x_{1}-x_{2})x_{3}\eta_{3}$ and
$D^{2}_{4}=(x_{1}+x_{2})r^{2}_{2}-(1-x_{1}-x_{2})x_{3}\eta_{3}$.

${\it 3.\ B^{+} \to \bar{D}^{*0}_{0} \pi^+\ decay:}$

In this decay, there are no annihilation contributions and  new
topologies involved. The corresponding flavor diagrams are the
same as Fig. \ref{ca_fig}(a) and (b) and Fig. \ref{cs_fig}. Hence,
we can immediately write the decay amplitude  as
\begin{eqnarray}
{\cal A}(B^+ \to \bar{D}^{*0}_{0} \pi^{+})=f_{\pi}
F_{BD^{*}_{0}}+M_{BD^{*}_{0}} + f_{D^{*}_{0}} F_{B\pi} + M_{B\pi}.
\end{eqnarray}
The hard amplitudes $\{ F\}$ and $\{ M\}$ are the same as Eqs.
(\ref{fe1}), (\ref{nonfe1}), (\ref{fe2}) and (\ref{nonfe2}).

\section{Numerical analysis \label{na}}

In our calculations, we adopt the $B$-meson wave function
$\Phi_{B}$ to be
\begin{eqnarray}
\Phi_{B}(x,b)&=& N_{B}x^{2}(1-x)^{2} \exp\Big[-\frac{1}{2}\Big(
\frac{x
\,m_{B}}{\omega_{B}}\Big)^{2}-\frac{\omega_{B}^{2}b^{2}}{2} \Big],
\end{eqnarray}
where $N_{B}$ can be determined by the normalization of the wave
function at $b=0$ and $\omega_{B}$ is the shape parameter. Since
the $\pi$-meson wave functions have been derived in the framework
of QCD sum rules, we display them up to twist-3 directly by
\cite{Ball}
\begin{eqnarray*}
\Phi_\pi(x)&=&\frac{3f_\pi}{\sqrt{2N_c}} x(1-x)
\left[1+0.44C_2^{3/2}(2x-1)+0.25C_4^{3/2}(2x-1)\right],
\\
\Phi_\pi^p(x)&=&\frac{f_\pi}{2\sqrt{2N_c}}
\left[1+0.43C_2^{1/2}(2x-1)+0.09C_4^{1/2}(2x-1)\right],\\
\Phi_\pi^{\sigma}(x)&=&\frac{f_\pi}{2\sqrt{2N_c}} (1-2x)
\left[1+0.55(10x^2-10x+1)\right],
\end{eqnarray*}
%%%%%%%%%%%%%%%%%%%
with the Gegenbauer polynomials,
\begin{eqnarray*}
C_2^{1/2}(\xi)&=&\frac{1}{2}(3\xi^2-1),\;\;\;\;\;
C_4^{1/2}(\xi)=\frac{1}{8}(35 \xi^4 -30 \xi^2 +3), \nonumber \\
C_2^{3/2}(\xi)&=&\frac{3}{2}(5\xi^2-1), \;\;\;\;\;
C_4^{3/2}(\xi)=\frac{15}{8}(21 \xi^4 -14 \xi^2 +1).
\end{eqnarray*}
After the wave functions of $\pi$ meson are determined, the
unknown $\omega_{B}$ can be fixed by decays such as $B\to \pi
\pi$. Consequently, the remaining uncertain parameters are the
wave functions of the $D^{*}_{0}$ meson.

To obtain the numerical results, the values of theoretical inputs
are chosen as: $\omega_{B}=0.4$, $f_{B}=0.19$, $f_{\pi}=0.13$,
$\tilde{f}_{D^{*}_{0}}=0.20$, $m_{B}=5.28$, $m_{D^{*}_{0}}=2.29$,
and $m^{0}_{\pi}=1.4$ GeV. With these values, we get the form
factor $F^{B\to \pi}(0)=0.3$.  In addition, the values of $B\to
D^{*}_{0}$ form factor with some variances in $a_{D^{*}_{0}}$ and
$\omega_{D^{*}_{0}}$ are also shown in Table
\ref{bsdff}. %%%%%%%%%%%%%%%%%%%%%%%%%%%%%%%%%%%%%%%%%%%%%%
%%%%%%%%%%%%%%%%%%%%%%%%%%%%%%%%%%%%%%%
\begin{table}[htb]
\caption{ The values of $B\to D^{*}_{0}$ form factor with
$b_{D^{*}_{0}}=0.5$ and some variances in  $a_{D^{*}_{0}}$ and
$\omega_{D^*_{0}}$.
 } \label{bsdff}
\begin{center}
\begin{tabular}{|c|c|c|c|}
% after \\: \hline or \cline{col1-col2} \cline{col3-col4} ...
\hline
 $\omega_{D^{*}_{0}}$ &  $a_{D^{*}_{0}}=0.7$ & $a_{D^{*}_{0}}=0.9$ & $a_{D^{*}_{0}}=1.1$ \\\hline
$0.5$ & $0.29$ & $0.30$  &  $0.31$\\ \hline $0.6$ & $0.24$ & $0.25$ & $0.26$ \\
\hline $0.7$ & $0.21$ & $0.22$ & $0.23$ \\\hline
\end{tabular}
\end{center}
\end{table}
%%%%%%%%%%%%%%%%%%%%%%%%%%%%%%%%%%%%%%%%%%%%%%%%%%%%%%%%%%%%
It is interesting that the form factor of $B\to D^{*}_{0}$ decay
is much smaller than that of $B\to D$ decay, which is calculated
to be around $0.57$ \cite{PQCD-Group}. We also find that our
results are a little bit larger than those calculated by ISGW2
model \cite{Cheng2}. According to Wolfenstein's parametrization
\cite{Wolfenstein}, we take $A=0.82$ and $\lambda=0.22$ for the
CKM matrix element $V_{cb}=A\lambda^{2}$. Hence, in terms of our
deriving formulas and by fixing $a_{D^{*}_{0}}=0.9$,
$b_{D^{*}_{0}}=0.5$ and $\omega_{D^*_{0}}=0.6$, the magnitudes of
the hard amplitudes are shown in Table \ref{va}.
%%%%%%%%%%%%%%%%%
\begin{table}[htb]
\caption{ The values of hard amplitudes (in units of $10^{-3}$)
with fixing $a_{D^{*}_{0}}=0.9$, $b_{D^{*}_{0}}=0.5$ and
$\omega_{D^*_{0}}=0.6$. } \label{va}
\begin{center}
\begin{tabular}{|c|c|c|c|c|c|}
% after \\: \hline or \cline{col1-col2} \cline{col3-col4} ...
\hline $f_{\pi} F_{BD^*_{0}}$ & $ M_{BD^*_{0}}$& $f_{B} F_{a}$ &
$M_{a}$ & $f_{D^*_{0}} F_{B\pi}$ & $M_{B\pi}$ \\\hline $36.4$ &
$10^{-2}(1.0-i 3.2) $ & $-0.06-i 0.08$ & $-1.85 -i 3.14$ & $-7.11$
& $7.56-i 10.95$ \\ \hline
\end{tabular}
\end{center}
\end{table}
 From the table, we can see clearly that except $M_{BD^{*}_{0}}$,
the nonfactorizable effects of color-suppressed process are
comparable to factorizable contributions; even in annihilation
topologies, the contributions of the former are much larger than
those of the latter. With fixing $b_{D^{*}_{0}}=0.5$ and
$\omega_{D^*_{0}}=0.6$ GeV and taking some different values of
$a_{D^{*}_{0}}$, the decay BRs of $B\to D^{*}_{0} \pi$  are
displayed in Table \ref{br1}. We also show the BRs with fixing
$a_{D^*_{0}}=0.9$ and $b_{D^*_{0}}=0.5$ and some variant values of
$\omega_{D^{*}_{0}}$. From both tables, we know that with proper
values of parameters, the calculated BR of $B^{+}\to D^{*0}_{0}
\pi^{+}$ is consistent with the BELLE's observation. It is worth
to note that the predicted BR of $B_{d}\to \bar{D}^{*0}_{0}
\pi^{0}$ is one order of magnitude smaller than others. The
phenomenon can be understood by noticing that, as shown in Table
\ref{va},
% as follows: according to Table \ref{va}, we see that
the value of $F_{B\pi}$ is very close and opposite in sign to the
real part of $M_{B\pi}$ such that there is a strong cancellation
in Eq. (\ref{nnamp}). As a result, we get the small BR in the
decay $B_{d}\to \bar{D}^{*0}_{0} \pi^{0}$. That is, the
annihilation effects are significant in $B\to \bar{D}^{*0}_{0}
\pi^{0}$.

%%%%%%%%%%%%%%%%%%%%%%%%%%%%%%%%%
\begin{table}[htb]
\caption{ The BRs (in units of $10^{-4}$) with fixing
$b_{D^{*}_{0}}=0.5$, $\omega_{D^*_{0}}=0.6$ GeV and various values
of $a_{D^{*}_{0}}$. } \label{br1}
\begin{center}
\begin{tabular}{|c|c|c|c|}
% after \\: \hline or \cline{col1-col2} \cline{col3-col4}
\hline $a_{D_{s}}$ & $B^{+}\to \bar{D}^{*0}_{0} \pi^{+}$ &
$B_{d}\to D^{*-}_{0} \pi^{+}$ & $B_{d}\to \bar{D}^{*0}_{0}
\pi^{0}$
\\\hline
  1.1 & 9.75 & 8.25 & 0.17  \\ \hline
  0.9 & 9.34 & 7.68 & 0.19  \\\hline
  0.7 & 8.98 & 7.13 & 0.21 \\\hline
\end{tabular}
\end{center}
\end{table}
%%%%%%%%%%%%%%%%%%%%%%%%%%%%%%%%%%%%%%%%%%%%%%
\begin{table}[htb]
\caption{ The BRs (in units of $10^{-4}$) with fixing
$a_{D^*_{0}}=0.9$ and $b_{D^*_{0}}=0.5$ and various values of
$\omega_{D^*_{0}}$. } \label{br2}
\begin{center}
\begin{tabular}{|c|c|c|c|}
% after \\: \hline or \cline{col1-col2} \cline{col3-col4}
\hline $\omega_{D^{*}_{0}}$ & $B^{+}\to \bar{D}^{*0}_{0} \pi^{+}$
& $B_{d}\to D^{*-}_{0} \pi^{+}$ & $B_{d}\to \bar{D}^{*0}_{0}
\pi^{0}$
\\\hline
 0.5 & 13.79 & 10.7 & 0.26  \\ \hline
 0.6 & 9.34 & 7.68 & 0.19  \\\hline
 0.7 & 6.28 & 5.55 & 0.15 \\ \hline
\end{tabular}
\end{center}
\end{table}

As stated before, although we only study the decays $B\to D^*_{0}
\pi$, we still can estimate the BRs of $B\to D^{(\prime)}_{1}
\pi$. Since
%Due to
only the longitudinal polarization has the contributions, except
the decay constants, we expect that the involving wave functions
of $D^{(\prime)}_{1}$ should be similar to $D^{*}_{0}$. By
neglecting the difference in phase space, the BRs of $B\to
D^{(\prime)}_{1} \pi$ could be estimated by $BR(B\to
D^{(\prime)}_{1} \pi)/BR(B\to D^{*}_{0}\pi) \sim
(\tilde{f}_{D^{(\prime)}_{1}} / \tilde{f}_{D^*_{0}})^{2}$. If
$\tilde{f}_{D^{(\prime)}_{1}} \approx \tilde{f}_{D^*_{0}}$, the
BRs for producing axial vector mesons $D^{(\prime)}_{1}$ are close
to that for the scalar $D^{*}_{0}$. The tendency is consistent
with BELLE's observations, shown in Eq. (\ref{belle-new}).

\section{Summary \label{summary}}

%By means of $D^{*}_{0}$,
We have studied the properties of $P$-wave mesons in $B$ decays in
terms of $D^{*}_{0}$. By taking the concept of the heavy quark
limit, we have obtained some information on the shapes of
$D^{*}_{0}$ wave functions. According to the wave function of
$K^*_{0}(1420)$, we can determine the proper value for the
parameter $a_{D^{*}_{0}}$ in $\Phi_{D1}(x)$. By the physical
argument, the unknown parameter $b_{D^*_{0}}$ can be chosen so
that the maximum of $\Phi_{D2}(x)$ locates at $x\sim 0.35$. We
have found that with a suitable value of $\omega_{D^*_{0}}$, our
result on $BR(B^{+}\to \bar{D}^{*0}_{0} \pi^{+})$ can fit BELLE's
measurements. Hence, the calculated BRs for $B_{d}\to D^{*-}_{0}
\pi^{+}$ and $B_{d} \to \bar{D}^{*0}_{0} \pi^{0}$ decays can be
viewed as our predictions. Finally, if we regard that
 the longitudinal wave functions of $D^{(\prime)}_{1}$ are the
same as $D^{(*)}_{0}$ and assume that
$\tilde{f}_{D^{(\prime)}_{1}} \approx \tilde{f}_{D^*_{0}}$, we
expect that the differences of BRs among them are not significant.
The more accurate predictions rely on  more definite values of
decay constants as well as other unknown
parameters.\\

 \noindent{\bf Acknowledgments}

The authors would like to thank C.Q. Geng, H.N. Li, H.Y. Cheng and
Taekoon Lee for their useful discussions. This work is supported
in part by the National Science Council of the Republic of China
under Grant No. NSC-91-2112-M-001-053 and the National Center for
Theoretical Sciences of R.O.C..\\

\end{document}